# On Nonlinear Waves in the Spatio-Temporal Dynamics of Interacting Populations

a) Ivan Jordanov, Nikolay K. Vitanov, Elena Nikolova,

b) Institute of Mechanics, Bulgarian Academy of Sciences, Sofia, Bulgaria

c) e-mail: i_jordanov@mail.bg, vitanov@imbm.bas.bg, elena@imbm.bas.bg

**Abstract**—In this paper the spatial-temporal dynamics of the members of interacting populations is described by nonlinear partial differential equations. We consider the migration as a diffusion process influenced by the changing values of the birth rates and coefficients of interaction between the populations. For the particular case of one population and one spatial dimension the general model is reduced to analytically tractable PDE with polynomial nonlinearity up to third order. By applying the modified method of simplest equation to the described model we obtain an analytical solution which describes nonlinear kink and solitary waves in the population dynamics.

## I. INTRODUCTION

The dynamics of many systems in biological, natural and social sciences is nonlinear [1]-[5]. If the nonlinearity compensates for the dispersion nonlinear traveling waves can propagate through the corresponding system. The dynamics of interacting populations is strongly nonlinear and the movement of humans or animals can lead to solitary migration waves. The solitary waves are very popular and frequently obtained in various areas of natural sciences [6]. Many authors have obtained exact solitary wave solutions which describe solitons, foldons, etc. in various equations such as Boussinesq equation [7], Korteveg-de Vries equation [8], sine-Gordon or sh-Gordon equations [9]-[13]. Below we will concentrate our attention on the analytical investigation of one-dimensional migration waves caused by members of one population. For the purpose a modified method of simplest equation proposed in the papers [14, 15] will be applied.

## II. MODIFIED METHOD OF SIMPLEST EQUATION

Let us briefly describe the modified method of simplest equation [14, 15], which is powerful tool for obtaining exact and approximate solutions of nonlinear PDEs. We have to solve a partial differential equation and let by means of an appropriate ansatz this equation be reduced to a nonlinear ordinary differential equation

$$P(F(\xi), \frac{dF}{d\xi}, \frac{d^2 F}{d\xi^2}, ...) = 0 \qquad (1)$$

For large class of equations from the kind (1) exact solution can be constructed as finite series

$$F(\xi) = \sum_{\mu=-\upsilon}^{\nu_1} P_\mu [\Phi(\xi)]^\mu \qquad (2)$$

where $\nu_1 > 0, \upsilon > 0, P_\mu$ are parameters and $\Phi(\xi)$ is a solution of some ordinary differential equation referred to as the simplest equation. The simplest equation is of lesser order than (1) and we know the general solution of the simplest equation or we know at least exact analytical particular solution(s) of the simplest equation. The modified method of simplest equation can be applied to equations of the kind

$$E(\frac{\partial^{\omega_1} F}{\partial x^{\omega_1}}, \frac{\partial^{\omega_2} F}{\partial t^{\omega_2}}, \frac{\partial^{\omega_3} F}{\partial x^{\omega_4} \partial t^{\omega_5}}) = G(F) \qquad (3)$$

where $\omega_3 = \omega_4 + \omega_5$. In the paper [14] the application of the modified method of simplest equation is based on the following steps:

1. By means of an appropriate ansatz (for an example the traveling-wave ansatz) the solved class of nonlinear PDE of kind (3) is reduced to a class of nonlinear ODEs of the kind (1).

2. The finite-series solution (2) is substituted in (1) and as a result a polynomial of $\Phi(\xi)$ is obtained. (2) is a solution of (3) if all coefficients of the obtained polynomial of $\Phi(\xi)$ are equal to 0.

3. By means of a balance equation one ensures that there are at least two terms in the coefficient of the highest power of $\Phi(\xi)$. The balance equation gives a relationship between the parameters of the solved class of equations and the parameters of the solution.

4. The application of the balance equation and the equalizing the coefficients of the polynomial of $\Phi(\xi)$ to 0 leads to a system of nonlinear relationships among the parameters of the solution and the parameters of the solved class of the equation.

5. Each solution of the obtained system of nonlinear algebraic equations leads to a solution a nonlinear PDE from the investigated class of nonlinear PDEs.

### III. APLICATION OF THE MODIFIED METHOD OF SIMPLEST EQUATION TO THE DYNAMICAL MODEL OF INTERACTING POPULATION

#### A. Formulation of the problem

We consider a two-dimensional plane populated by members of $n$ populations. We will investigate the density $\rho_i$ of the $i$-th population. We assume that that the members of the $i$-th population can move through the borders of a small area under consideration. Then we can define the density $I_i$ of the current of this movement. Within the observed area the number of the members of the $i$-th population can change by births, deaths or interactions with other populations. These processes will be summarized mathematically by the function $R_i$ below. In this way the total change of the number of members from the $i$-th population in some large area is a consequence of migration through the boundaries of the area and the birth, death and interaction processes within the area. As a mathematical expression this reads

$$\frac{\partial \rho_i}{\partial t} + \nabla \cdot I_i = R_i, \quad i = 1, 2, \ldots, n \quad (4)$$

The function $R_i$ can be taken in the form [16]:

$$R_i = r_i^0 \rho_i [1 - \sum_{j=1}^{n} (\alpha_{ij}^0 - r_{ij}) \rho_j - \sum_{j,k=1}^{n} (\alpha_{ijk}^0 + r_{ik}) \rho_j \rho_k - \sum_{j,k,l=1}^{n} \alpha_{ij}^0 r_{ik} \alpha_{ijl} \rho_j \rho_k \rho_l] \quad (5)$$

Eq.(5) is obtained as follows. As in the case of the generalized Lotka - Volterra equations we assume that the changes of the density because of reproduction and interaction between the populations are

$$R_i = r_i \rho_i [1 - \sum_{j=1}^{n} \alpha_{ij} \rho_j] \quad (6)$$

where $r_i$ is the growth ratio of the $i$-th population and $\alpha_{ij}$ is the interaction coefficient measuring to what extent the growth of the $i$-th population is influenced by the $j$-th population. We assume that the both coefficients are density dependent and that they depend on the density of the members of the populations in the following manner:

$$r_i = r_i^0 [1 + \sum_{k=1}^{n} r_{ik} \rho_k]$$

$$\alpha_{ij} = \alpha_{ij}^0 [1 + \sum_{l=1}^{n} \alpha_{ijl} \rho_l] \quad (7)$$

In addition we assume that $I_i$ has a form similar to the general form of the linear multi-component diffusion.

$$I_i = -\sum_{j=1}^{n} D_{ij} \nabla \rho_i \quad (8)$$

For the case of one population and one spatial dimension (4) reduces to the following model equation

$$\frac{\partial \rho}{\partial t} - D \Delta \rho = r^0 \rho [1 - (\alpha^0 - r_{11}) \rho - \alpha^0 r_{11} \rho^2] \quad (9)$$

The fixed points of (9) correspond to stationary densities $\rho_0$ and can be found by solving the equation

$$\alpha^0 r_{11} \rho_0^2 + (\alpha^0 - r_{11})\rho_0 - 1 = 0 \quad (10)$$

We are interesting in the behavior of deviation around the basic state $\rho_0$, i.e.

$$\rho(x,t) = Q(x,t) + \rho_0 \quad (11)$$

The substitution of (11) in (9) leads to the following nonlinear equation for $Q$

$$\frac{\partial Q}{\partial t} - D\frac{\partial^2 Q}{\partial x^2} = FQ^3 + GQ^2 + HQ \quad (12)$$

where

$$F = -r^0 \alpha^0 r_{11};$$
$$G = r^0(r_{11} - \alpha^0 - 3\rho_0 r_{11} \alpha^0) \quad (13)$$
$$H = r^0[1 - \rho_0(3\alpha^0 r_{11}\rho_0 + 2r_{11} - 2\alpha^0)]$$

Below we will treat (12) as a general form of a model equation which for the particular case when $F$, $G$, $H$ satisfy the above relationships describes population density waves.

### B. Traveling wave solutions of (12)

We consider the equation (12) and search for solutions of the kind of traveling waves: $Q(x,t) = Q(\xi) = Q(x-vt)$, where $v$ is the velocity of the wave. We have to solve the equation

$$D\frac{\partial^2 Q}{\partial \xi^2} + v\frac{dQ}{d\xi} + FQ^3 + CQ^2 + HQ = 0 \quad (14)$$

We assume that $Q(\xi)$ has the form

$$Q(\xi) = \sum_{i=0}^{n} a_i \phi^i, \quad \left(\frac{d\phi}{d\xi}\right)^2 = \sum_{j=0}^{r} c_j \phi^j, \quad (15)$$

where $a_i$ and $c_j$ are parameters that we will determine below. Following the steps of the modified method of the simplest equation we substitute (15) in (14) and obtain an equation that contains powers of $\phi$. Next we balance the highest power arising from the second derivative in (14) with the highest power arising in the term containing $Q^3$ in the same equation. The resulting balance equation is

$$r = 2n+2, \quad n = 2,3,... \quad (16)$$

If $n=2$ then $r=6$. Assuming that $c_0 = c_1 = c_3 = c_5 = 0$, $c_2 = p^2, c_4 = 2pq, c_6 = q^2 \neq 0$ we search for a solution of the form

$$Q(\xi) = a_0 + a_1\phi + a_2\phi^2,$$
$$a_2 \neq 0, \quad \frac{d\phi}{d\xi} = p\phi + q\phi^3 \quad (17)$$

Substituting this in (14) we obtain the following system of 7 algebraic equations

$$8Dq^2 + Fa_2^2 = 0$$
$$Fa_1 a_2^2 + Da_1 q^2 = 0$$
$$3Fa_2 a_0 + 3Fa_1^2 + Ga_2 + 2vq + 12Dpq = 0$$
$$4Dpq + 6Fa_0 a_2 + Fa_1^2 + vq + 2Ga_2 = 0$$
$$3Fa_0^2 a_2 + 3Fa_0 a_1^2 + 4Dp^2 a_2 2va_2 p +$$
$$+ Ha_2 + 2Ga_0 a_2 + Ga_1^2 = 0$$
$$3Fa_0^2 a_1 + Dp^2 a_1 + 2Ga_1 + va_1 p + H = 0$$
$$Ga_0 + Ha_0^2 + Fa_0^3 = 0 \quad (18)$$

The system (18) implies that $a_1 = 0$. The solution of (18) is:

$$a_2 = \frac{\pm 2q\sqrt{-2FD}}{F};$$
$$a_0 = -\frac{G\sqrt{-2FD} + vF + 6DpF}{3F\sqrt{-2FD}};$$
$$H = \frac{G^2 D + v^2 F + 6D^2 p^2 F}{FD}; \quad (19)$$
$$p = -\frac{2GF^2 Dv - 2\sqrt{-2FD}v^2 F^2 + G^{2\,3/2}\sqrt{-2FD}}{12F^2 D(GD + v\sqrt{-2FD})}$$

The expression for the solitary wave depends on the solution of the differential equation in (17) and it is given by

$$Q(\xi) = a_0 + a_2 \sqrt{\frac{p \exp[2p(\xi+c)]}{1 - q \exp[2p(\xi+c)]}} \quad (20)$$

where $a_0, a_2, p$ are given by (19). (20) describes a kink wave (an example is shown in Fig. 1). This kink can be interpreted as the front of the migration of the studied population.

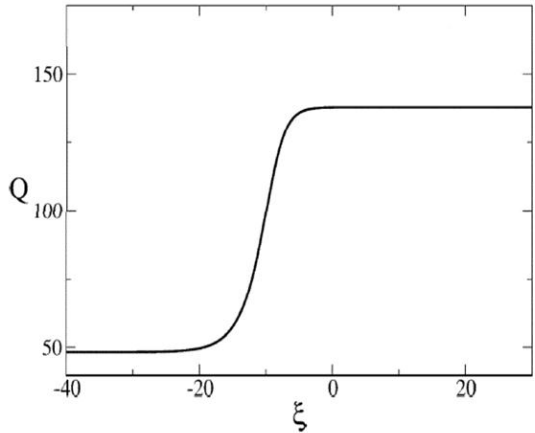

Figure 1. Graph of the solution $Q(\xi)$ at $\alpha_{111} = 0; D = 0.1; G = -0.1; r_0 = 100; \alpha_0 = 0.02; \rho_0 = 50; q = -1000$

## IV. Conclusions

In this paper we discuss a nonlinear model of the spatial-temporal dynamics of interacting populations. We consider in more details the particular case of 1D migration of 1 population. For this case the model system is reduced to a single (1+1)-dimensional nonlinear PDE for which an exact analytical solution can be obtained. This equation describes the evolution of the density deviation from the states of stationary spatial density of the population members. If the amplitude of such a deviation becomes large solitary waves can travel through the system. In addition by means of appropriate ansatz we obtain an exact particular analytical solution of the model equation. This solution describes nonlinear kink and solitary wave, expressing the spreading of the population density changes in the space. Our further aim is to generalize the same theory for the case of two populations. In this case the phenomenon of coupled kink waves (traveling with the same velocity) will be observed.